%% file: main.tex
\documentclass[submission,copyright,creativecommons]{eptcs}
\providecommand{\event}{AppFM 2021} 
\def\titlerunning{Teaching Code Analysis and Verification}
\def\authorrunning{S. Souaf et al.}
\input{preambule}

\title{Experience Report: Teaching Code Analysis and Verification Using Frama-C}
\author{Salwa Souaf 
\institute{CentraleSupélec\\
Rennes, France}
\email{salwa.souaf@centralesupelec.fr}
\and
Frédéric Loulergue
\institute{
  Univ Orléans, INSA CVL, LIFO EA 4022\\
  Orléans, France}
\email{frederic.loulergue@univ-orleans.fr}
}
\def\titlerunning{Teaching Code Analysis and Verification Using Frama-C}
\def\authorrunning{S. Souaf and F. Loulergue}

\begin{document}
\maketitle

\begin{abstract}
\input{abstract.tex}

\end{abstract}

\section{Introduction}
\input{introduction}

\section{Context and Background}
\label{sec:context}
\input{context}

\section{Teaching Approach}
\label{sec:approach}
\input{approach}

\section{Students Assessment and Students Evaluation}
\label{sec:students}
\input{students}

\section{Conclusion and Related Work}
\label{sec:conclusion}
\input{conclusion}

\bibliographystyle{eptcs}
\bibliography{biblio}

\end{document}

%% file: preambule.tex
\usepackage[utf8]{inputenc}
\usepackage{graphicx}
\usepackage{enumitem}
\usepackage{todonotes}

\usetikzlibrary{decorations.pathreplacing,trees,arrows}
\usetikzlibrary{matrix}
\usetikzlibrary{shapes}
\usetikzlibrary{calc}
\usetikzlibrary{fit}
\usetikzlibrary{decorations.text}
\usetikzlibrary{positioning}
\tikzset{markplace/.style=
 {rectangle callout, fill=orange!40,
  callout absolute pointer={#1},
  at={#1}, above=1cm 
 }}

%% file: abstract.tex
Formal methods provide systematic and rigorous techniques for software development. We strongly believe that they must be taught in computer science curricula. In this paper we present the pedagogic rationale and the concrete implementation of two courses on the use of formal methods, sharing some material. These courses promote the usage of formal verification to ensure safety and security of software, exemplified in the domain of the Internet of Things.

%% file: introduction.tex
To make formal methods more widely applied, one way is to make them
easier to use, in particular by making them more automatic and
seamlessly integrated in software engineering tools and
processes. However, full automation for deductive verification is not
possible and we think it is desirable that more professionals know it
is possible to formally specify and verify real-world programs, and
have a hands-on experience to do it. Teaching program verification is
a way to contribute to this goal.

This paper reports teaching experiences on C programs verification
using Frama-C~\cite{FramaC,BBB2021:CACM}, an open-source framework for
the analysis and verification of ANSI C99 programs.  The approach is
to present the concepts mainly through examples and have the students
specify and verify real-world code. Sharing some material and a
project, we taught two courses in different institutions: Northern
Arizona University, USA and INSA Centre Val de Loire, France. Over the
years efforts of teaching and integrating formal methods in software
engineering courses faced many challenges, some of which are mentioned
in~\cite{2016:ENSASE:Spichkova}. We encountered several of these
challenges, but in this paper, we rather describe the more focused and
detailed difficulties our students faced.

The paper is organized as follows: Section~\ref{sec:context} describes
the context including the Frama-C tool and the position of the courses
in their respective programs, as well as our
audiences. Section~\ref{sec:approach} is devoted to the approach we
took for these courses: the organization, content and assignments, in
particular the project. Students' challenges, results and evaluations
are given in Section~\ref{sec:students}. We conclude and mention
related work in Section~\ref{sec:conclusion}.

%% file: context.tex
\begin{figure}
  \centering
  \includegraphics[width=.9\textwidth]{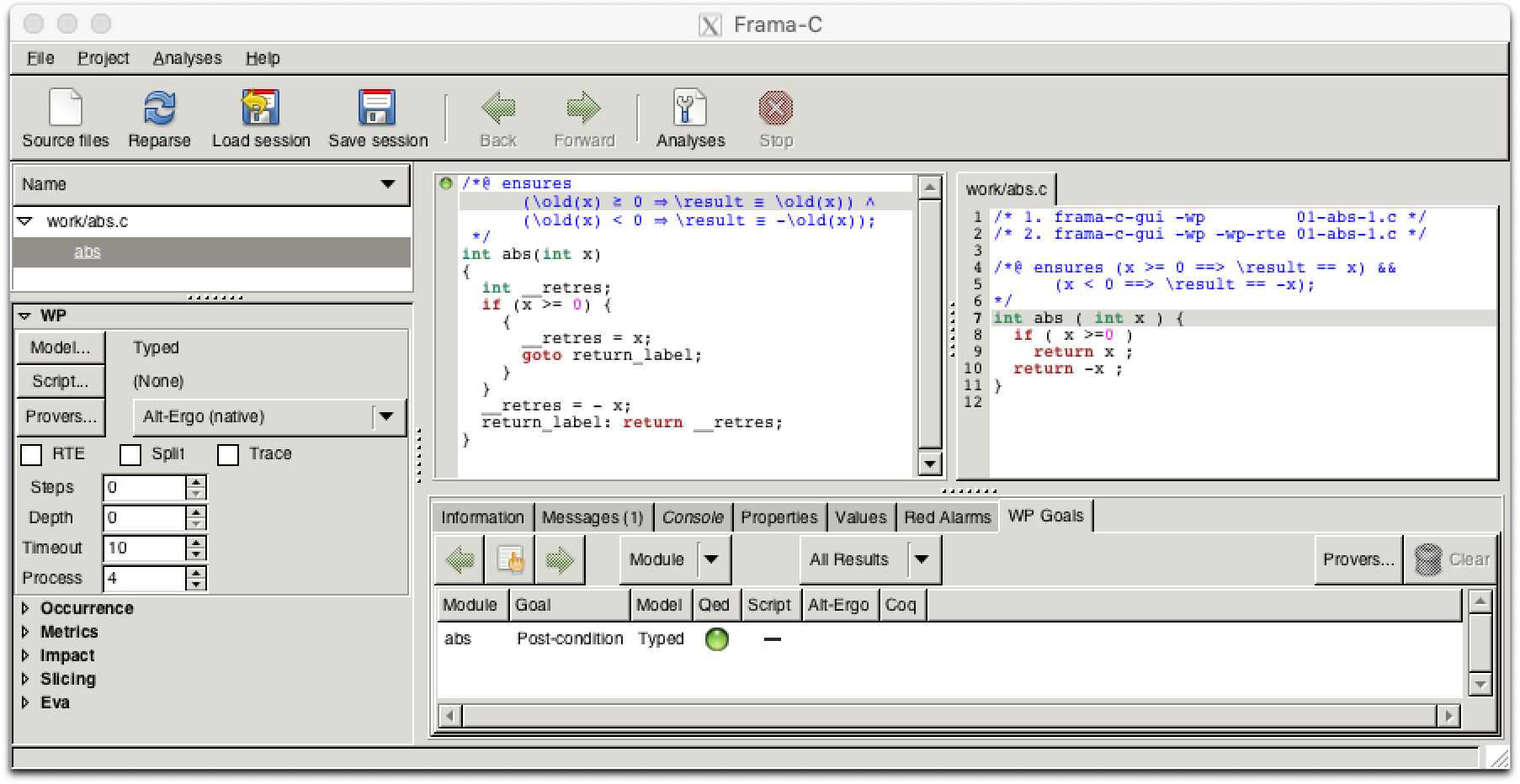}
  \caption{An example of use of Frama-C}
  \label{fig:example}
\end{figure}

Frama-C is architectured as a set of plugins built around a kernel that provides basic services (such as parsing and access to ASTs of the programs). Plugins use, generate, or verify annotations written in the ACSL (ANSI C Specification Language). Analyses and verification techniques include (but are not limited to) value analysis by abstract interpretation (plugin EVA), deductive verification (plugin WP) and dynamic verification (plugin E-ACSL). The plugin RTE is dedicated to generate ACSL assertions for each expression potentially leading to an undefined behavior. Frama-C is used both in industry and academia. The courses presented in this paper evolved from a series of tutorials given at various international conferences (in particular~\cite{BKL2018:HPCS,BKL2018:SECDEV}) and experience from research related to Frama-C (for example~\cite{BKL2016:SCAM,BKL2019:SAC,BLK2019:NFM}). Figure~\ref{fig:example} presents a screenshot of the tool while working on one of the first examples of the class with the WP plugin. On the right panel, the original code with an ASCL contract (only a post-condition) on lines 4--5 is displayed. The way the code is considered by the tool as well as a pretty-printed version of the contract is displayed on the left panel. The green bullet indicates that the function is correct with respect to this contract. Indeed the tool is used here without the RTE plugin, therefore there is not any assertion in the code that checks whether there is an overflow when evaluating the expression \texttt{-x}. If the RTE plugin is used, such an assertion is generated and the bullet is orange meaning the tool fails to verify that the code respects the contract.

At Northern Arizona University (NAU), the course described in this paper is course CS~451 Mechanized Reasoning about Programs. The only prerequisite were some standard architecture and data structure classes. The focus was on applied formal methods, and we chose to build it around Frama-C. The version of the class described here was taught in Fall 2019 to 37 bachelor students, most of them in their fourth year in the Computer Science program or the Applied Computer Science program. At INSA (Institut National des Sciences Appliquées) Centre Val de Loire, this course was taught to students in their fourth year. INSA is a network of French engineering schools. Each school offers five year programs from which students graduate with a Master's degree. The students who followed this course majored in Security and Computer Technologies. These students have a strong background in programming using different languages but were not exposed to theoretical courses covering formal methods. This course was taught to 23 students during Spring 2020.

%% file: approach.tex
At NAU, CS~451 was 2.5 hours a week, for 15 weeks. In-class exercises
were given to students after each concept and related examples were
presented, some of which were group exercises. In addition to 8
homework assignements, students had to complete a project in three
phases. Finally, a midterm and a final exams were organized.  For the
homework assignments, in addition to the keys, extensive feedback was
presented in class: the most common errors were discussed, as well as
the various possible ways to correctly solve the exercise and their
respective merit. All the assignments needed to write ACSL annotations
and use Frama-C. All the material presented in class was available
online, as well as keys for the homework assignments, midterm, and
project. A mock midterm exam was organized a week before the exam. No
textbook was mandatory for this course. However, the students were
referred to Allan Blanchard's tutorial~\cite{BLA2019:ZDS} and Jens
Gerlach et al. ACSL by Example~\cite{ACSLbyExample}.

For the fourth year students at INSA Centre Val de Loire, the course was divided into 10h40 for the lectures and 10h40 for the in-class work. In order to fully utilize this short time and deliver the course content in a more digestible approach, the decision was made to give the lectures in a tutorial like style to introduce Frama-C and the in-class work as an implementation project with each session devote to a different phases of the project. The students were assessed based on their project work.

\paragraph{Northern Arizona University}

The course was initially divided into four parts: 6 weeks devoted to the course introduction, basic concepts of ACSL and deductive verification with Frama-C/WP (and RTE), 2 weeks for the foundational aspects: axiomatic semantics (and the reference for this part was~\cite{NN1992:WILEY}), 5 weeks for additional concepts of ACSL and deductive verification with application on case studies and finally 2 weeks for an introduction to static analysis with Frama-C/EVA. The three first parts took longer than planned, and the fourth part was removed. The second and third parts were swapped.

ACSL is basically first-order logic with typed C expressions, typed logical expressions and specific predicates. After presented Frama-C as a whole and basic usage of the tool, the course presented the concept of function contract with pre- and post-conditions. The first contracts contained only non-quantified annotations, and without specific predicates.
In ACSL annotations mathematical types are used: a C variable containing an integer is thus considered as a mathematical integer. The plugin RTE adds assertions into the code to verify there is not any runtime error. This includes for example signed integer overflows. The use of RTE was introduced at this point in the course. Then the \texttt{\textbackslash{old}} predicate, to be used in post-conditions, was introduced. Finally, the \texttt{assign} operator, which indicates which parts of the memory are potentially modified was explained. Note that up to this point, pointers were not used in the examples and exercises. How contracts are used in function calls was included in this introductory presentation. This was completed by the presentation of ACSL behaviors which allow to define contracts for sub-parts of the input space, the  pointer-specific predicate \texttt{\textbackslash{valid}}, and for expressing properties on arrays, quantifiers. This introductory part ended with set expressions in ACSL, to express memory validity more concisely. The final example was the specification, using behaviors, of a \texttt{find} function on an array of integers, that returns the index of a searched value, or \texttt{-1}. Contrary to previous examples and exercises, Frama-C/WP is unable to check the correctness of the implementation with respect to this contract: loop annotations are necessary. The first part of the class ended with all the concepts related to loop annotations: invariants, variants, loop assigns. Loop annotations were related to mathematical induction, and a significant part was devoted to methods to find invariants.  

The second part introduced additional ACSL concepts in order to make the project doable: labels, user-defined predicates, the statement of lemmas, logic functions, and finally ghost code.

The goal of the third part was to present axiomatic semantics in a formal way and weakest precondition calculus in an informal way (as a method to prove axiomatic semantics judgments). It however started with the informal presentation of a small imperative language called nano-C (similar to the usual IMP or While languages but with a C syntax) and its structural operational semantics, with a small modification with respect to IMP/While: sequence is not a command in nano-C, but there are lists of commands (the main program, branchs of conditionals, loop body). This makes the rules of the operational semantics all axioms.

\paragraph{INSA Centre Val de Loire}

Unlike the version of the course described above, the time dedicated in INSA did not allow an application of a classic detailed presentation of the different ACSL concepts, deductive verification and static analysis. For this specific reason and knowing the background of the students, we took a tutorial-style approach. The idea was to present the different concepts of ACSL and steps of code specification by relying on a concrete small example presented and tested live by the instructor, then  a different example for the students to do in class and corrected during the same session. This was a way for the students to practice using the Frama-C tools from the very first sessions, get more familiar with the ACSL specification language as well as keep them invested and attentive throughout the session. We focused on utilizing and presenting, in this order, the RTE, WP and EVA plug-ins as well.

\paragraph{Project: Contiki's memb module}
\input{project}

%% file: project.tex
As demonstration of the effectiveness and usability of code specification and verification, we proposed a project that aims to specify and verify the \texttt{memb} module of the Contiki OS. The subject of this project was inspired from the work presented in~\cite{2016:CRISIS:Mangano}. 

Contiki is an operating system for networked, memory-constrained systems with a focus on low-power wireless Internet of Things devices. Uses for Contiki include systems for street lighting, sound monitoring for smart cities, radiation monitoring, and alarms. It is open-source software. Contiki was developed in 2002, the main focus was on enabling communication in the most constrained devices, with no particular attention given to security. As it matured and as commercial applications arose, communication security was added at different layers, via standard protocols such as IPsec or DTLS. The security of the software itself, however, did not receive much attention.

The \texttt{memb} module is Contiki’s main memory management module. To avoid fragmentation in long-lasting systems, Contiki does not use dynamic allocation. Memory is pre-allocated in blocks on a per-feature basis, and the \texttt{memb} module helps the management of such blocks. It offers a simple API, enabling to initialize a \texttt{memb} store, to allocate a block, to free a block, to check if a pointer refers to a block inside the store and to count the number of allocated blocks. The \texttt{memb.c} file consists in about 100 lines of C code but is one of the most critical elements of Contiki, as the kernel and many modules rely on it. The Contiki code base involves a total of 56 instances of \texttt{memb}. Not all are included in a given Contiki firmware, but a subset is included depending on the application and configuration. \texttt{memb} is used for instance for HTTP, CoAP, IPv6 routes, the MAC protocol TSCH, packet queues, etc.

The goal of the project was to formally specify -- in ACSL -- and verify -- using Frama-C/WP -- the \texttt{memb} module. The API studied in this project is extremely close to the actual Contiki-NG API, but the provided implementation differs for some functions, to make the verification easier for students. Elements to guide them through the different phases were detailed in the document they received at the beginning of each phase. The files that were provided to students were: \textit{logic\_defs.h} containing user-defined predicates and logic functions; \textit{lemmas.h} to help the provers to verify the functions with respect to their contracts; \textit{memb.h} headers of the \texttt{memb} API; and \textit{memb.c} the full implementation.

The project was divided into three phases in order to better structure the work and have tangible intermediate deliverables. This also eased the amount of work for students and helped them get organized. When faced with a global project description without some guidance, they tend to either be overwhelmed, not knowing where to start, or to push it off to the last minute.

\noindent\textbf{Phase~1} students had to write several user-defined predicates to be able to make function contracts more concise and readable. These predicates were used to describe properties of \texttt{memb} stores.  The two functions to specify and verify (including termination) were: \texttt{memb\_init} and \texttt{memb\_inmemb}.
   
\noindent\textbf{Phase~2} focused on the specification and verification (including termination) of two functions of the API: \texttt{memb\_numfree}, which returns the number of free blocks in the memory store, and  \texttt{memb\_free}, which frees a block, if possible, i.e. if the pointer to free is in the memory store, if it is aligned (c.f. predicate memb aligned), and if it is not already free.

\noindent\textbf{Phase~3} focused on the last function of the API: \texttt{memb\_alloc}, which returns the address of the first free block if the memory store is not full, and \texttt{NULL} otherwise. Unlike the functions of the previous phase, to verify \texttt{memb\_alloc}, some assertions, in addition to the loop annotations, were needed in the body of the function for the automatic solvers to verify the contract. Some of them were provided, but not all.

%% file: students.tex
\paragraph{Students assessment}

At NAU, at the very beginning of the class, a reminder of propositional calculus truth tables was necessary for some students. On one hand, some students were not acquainted with mathematical notations for logical connectors: ASCL code is written using C syntax for common operations, for example \texttt{\&\&} for conjunction, or specific ACSL syntax, for example \texttt{==>} for implication, but by default Frama-C pretty prints these symbols as $\wedge$ and $\Rightarrow$ which was confusing to some students. In the other hand, when presenting the ACSL \texttt{\textbackslash{valid}} predicate, some students were not comfortable with C-specific features such a pointers, because they essentially programmed using language with automated memory management. In the class CS~396 Principles of Programming Languages, parameter-passing modes are presented. But a significant subset of the students had not taken this class yet, so it was actually necessary also to take some time for C reminders about pointers but also parameter passing.

For an array of size \texttt{n}, two usual ACSL patterns are \texttt{\textbackslash{forall} integer k; 0<= k < n ==> P[k]} and \texttt{\textbackslash{exists} integer k; 0<= k < n \&\& P[k]} to express that all cells of the array satisfy a property \texttt{P} and there exists a cell that satisfies a property \texttt{P}. Understanding that using \texttt{\&\&} instead of \texttt{==>} in the universal quantification case and vice-versa in the existential case does not work was a challenge for a significant number of students. Most of the students who failed the class were lost at this point. 

Loop invariants are of course challenging. We encouraged the students to write loop annotations incrementally. First the loop assigns annotation and simple invariants about the loop counter when there is one. Most students were able to do so. Writing annotations needed to prove post-conditions was of course more difficult. Related to the challenged described in the previous paragraph, in the \texttt{find} example (and similar cases) writing a loop invariant that is the negation of an existential in the post-condition was a problem. Making the students explain informally why the loop runs rather than stops helped. 

The final grades are given in the following table. 76\% of the enrolled students passed the course. 
\begin{center}
  \begin{tabular}{|c|c|c|c|c|c|}
    \hline
    \textbf{Grade} & A  & B & C & D & F \\
    \hline
    \textbf{Count} & 13 & 8 & 7 & 2 & 7 \\
    \hline
  \end{tabular}
\end{center}

All the passing students were able to read and write a function contract including behaviors, and user-defined predicates. They were also able to write loop annotations, but students with a C grade had difficulties in obtaining a strong enough invariant to allow the proof of post-conditions and some form of logical functions may be challenging. Students with a B grade had problems with writing additional annotations to guide the SMT solvers, when students with a C grade were not able to write such annotations. While all passing students were able to write correct contracts, and were able to check whether pre-conditions contain a logical contradiction, they had sometimes difficulties to write strong enough contracts of called functions to allow the verification of calling functions.

At INSA, the students were assessed based on their final work in the project. The first phase was a test phase to make sure that the majority of the students understood the task at hand and have a good comprehension of the different aspects presented during the class. The second and third phases were the ones taken into consideration in their final grade. The following table summarizes the students grades.

\begin{center}
  \begin{tabular}{|c|c|c|c|c|c|}
    \hline
    \textbf{Grade (over 20)} & $>16$ & 15--16  & 13--14 & 10--12 & $<10$ \\
    \hline
    \textbf{Count} & 4 & 3 & 6 & 4 & 6  \\
    \hline
  \end{tabular}
\end{center}

While most of the students were able to write function contacts as well as loop annotations during class for smaller examples, when it came to the project they had difficulties executing. The students with grades less than 12 had difficulties in writing loop annotations specially with in finding an invariant. Students with grades between 13 and 15 had a good understanding of the level of specification needed but failed to find exactly the key annotations to guide the solvers. Finally, some students have shown quite the potential, they were able in a short amount of time to well execute the knowledge acquired in the class.

\paragraph{Students evaluation}

At NAU, there is a procedure for class evaluation by students: 7 questions with 4 possible responses ranging from strongly disagree (1) to strongly agree (4), and 3 open questions: What suggestions do you have to improve this course? The assignment that most contributed to my learning is. What did you like best about this course?
The average score to the 7 questions was 3.5 which denotes a good level of satisfaction. There was not any heavy trend in the answers to open questions, but for the assignment that helped the most: the homework assignments. This is not surprising as these assignments are focused on a few concepts. In-class group exercises were also appreciated.

Suggestions for improvements were often contradictory, for example one student would have liked even more examples while another one would have prefered a focus on concepts. Some comments focused on the last assignment but made it a general comment on the course: in the last phase of the project, the students had to experiment with writing assertions in a function body to guide the SMT solvers. Thwas is not an aspect we exercised a lot in class. For all other concepts and ACSL features, several examples and exercises were presented. One comment was representative of the challenges faces by some students: ``have an assignment that explicitly outlines/teaches mathematical logic such as disjunction, conjunction, proof by induction, and the like.'' 

There were many aspects the students liked best about this course. The only answer with several (similar) occurrences was: ``thinking about the code in a different way''.

%% file: conclusion.tex
In~\cite{2019:FMTea:Leo}, Creuse et al. present a course on formal methods that focuses on deductive verification through Frama-C and SPARK. This course was aimed at students with no strong background in computer science. In their paper, Creuse et al. detail two experiments made at ISAESUPAERO in an engineering program focusing on aerospace industry. Blazy presents an initiation course to formal methods in~\cite{2019:FMTea:Blazy}. She presents her approach and methodology in teaching deductive verification using the Why3 platform. This course was aimed at third-year students at the University of Rennes 1 in France.

Our, rather successful, approach is a hands-on approach using Frama-C, for fourth-year students. In addition to the important concepts of deductive verification, pre- and post-conditions, loop invariants and variants, the students experienced by themselves that it is possible in practice to formally specify and verify real-world C programs using a tool.